\documentclass[12pt]{iopart}
\usepackage{graphicx}
\usepackage{iopams}  

\renewcommand{\d}{\rmd}

\newcommand{\text}[1]{\mathrm{#1}}
\newcommand{\kt}{k_\perp}
\newcommand{\qt}{q_\perp}

\newcommand{\alphas}{\alpha_\mathrm{s}}

\newcommand{\qhat}{\hat q}

\begin{document}

\title{Monte Carlo Tools for Jet Quenching}

\author{Korinna Zapp}

\address{Institute for Particle Physics Phenomenology, Durham University, \\ Durham\ 
DH1\,3LE, UK}
\ead{k.c.zapp@durham.ac.uk}
\begin{abstract}
 A thorough understanding of jet quenching on the basis of multi-particle final
states and jet observables requires new theoretical tools. This talk summarises
the status and propects of the theoretical description of jet quenching in
terms of Monte Carlo generators.
\end{abstract}


\section{Introduction}

The motivations for studying jet observables instead of single-inclusive
observables are manifold and shall not be discussed extensively here. Instead,
only two important points will be mentioned. Firstly, it has turned out that
single-inclusive observables do not fully constrain the analytical models for
partonic energy loss. The nuclear modification factor for instance is
described by all models equally well albeit with very different transport
coefficients\cite{Bass:2008rv}. The benefit of studying sub-leading fragments
and jet
observables is that these are much more discriminating. The downside is that
they are not well modelled by the existing analytical calculations and new
theoretical tools are needed.

Secondly, on the experimental side there is need for reliable tools that allow
to disentangle jets from the background. This requires a quantitative
understanding of jet areas, background and background fluctuations\cite{Cacciari:2011tm} 
and the response of jet finding algorithms to quenched jets\cite{Bruna:2010sz}.
Further complications arise from the fact that the implicit assumption that jets
and background
are uncorrelated is strictly speaking not justified. It is to be expected on
general grounds that there is a backreaction of the jets to the medium, but this
is very difficult to quantify with the currently available tools. It is,
however, clear that a quantitative understanding of jets, background and their
correlations requires running jet finders on the theory predictions. 

\smallskip

In particle physics Monte Carlo (MC) event generators have emerged as extremely
powerful and versatile tools linking theory and experiment. It is the aim of
this talk to make an assessment of the status of MC models for jet quenching by
highlighting conceptual issues and to what extent they are solved or solvable
using MC tools. It focusses on dedicated jet quenching MC models and does not
discuss parton cascade codes. The latter should in principle also be capable of
describing jet quenching, but as the approach and focus are quite different from 
specialised jet quenching MC codes they will not be dicussed here.

\section{Jets in p+p}
\label{sec_vacjets}

Jet evolution in vacuum arises due to collinear divergences in real emission
matrix elements. The singularity structure is universal, leading to a
factorisation in the collinear region. The differential
cross section for a process with $n+1$ partons can thus be written as the
product of the
differential cross section for the process with $n$ partons, the single
particle phase space and the differential radiation probability characterised
by the Altarelli-Parisi splitting function:
\begin{equation}
 \d \sigma_{n+1} \approx \d \sigma_n \frac{\d t}{t}\, \frac{\d \phi}{2 \pi}\, \d
z\, \frac{\alphas}{2\pi} \mathcal{P}(z)
\end{equation}
The variable $t$ quantifies the hardness of the splitting and can be the
transverse momentum squared $\kt^2$, the virtuality $Q^2$ or the radiation angle
squared $\vartheta^2$ -- all these choices are to leading logarithmic accuracy
equivalent. The regularisation of the $1/t$-singularity gives rise to large
logarithms that can compensate the smallness of the strong coupling $\alphas$
and thus have to be resummed to all orders. This proceeds via a DGLAP evolution
equation, which for only one parton species can be written as
\begin{equation}
 t \frac{\partial}{\partial t} f(x,t) = \int\limits_x^1 \d z\,
\frac{\alphas}{2 \pi} \mathcal{P}(z) \left[ \frac{1}{z} f(x/z,t) -
f(x,t) \right] \,.
\end{equation}
Here, $f(x,t)$ is the parton density at energy fraction $x$ and scale $t$. The
change in $f(x,t)$ when increasing the scale is given by an increase due to
splittings of partons at higher energy fraction $x$ and a decrease due to
partons at $x$ that split to populate smaller energies. This equation can be
integrated with the help of the Sudakov form factor defined as 
\begin{equation}
 \mathcal{S}(t_1,t_2) = \exp \left\{ - \int\limits_{t_1}^{t_2} \frac{\d t}{t}
\int \d z\, \frac{\alphas}{2 \pi} \mathcal{P}(z) \right\} \,.
\end{equation}
The integrated evolution equation then becomes
\begin{equation}
 f(x,t) = \mathcal{S}(t_0,t) f(x,t_0) + \int\limits_{t_0}^t
	  \frac{\d t'}{t'} \mathcal{S}(t',t) \int \frac{\d z}{z} 
\frac{\alphas}{2\pi} \mathcal{P}(z) f(x/z,t') \,,
\end{equation}
indicating that $\mathcal{S}(t_0,t)$ can be interpreted
as the probability for having no emission between $t$ and $t_0$. Here, $t_0$ is
the
infra-red cut-off scale, at which the dynamics becomes dominated by
non-perturbative physics. As it regularises a soft and collinear divergence it
is important that the observables that are to be calculated are insensitive to
the exact value of the cut-off, i.e.\ are infra-red and collinear safe.

The Sudakov form factor is the basis for an iterative Monte Carlo (MC)
algorithm that generates the radiative corrections to any given hard process and
is called the parton shower. It resums real emissions to all orders to leading
logrithmic accuracy and via unitarity also includes the corresponding virtual
corrections.

\medskip

The elements of Monte Carlo event generators that are relevant for jet
production and evolution are matrix elements, parton showers and the
hadronisation. The matrix elements describe hard scattering processes and are
calculated at fixed order in perturbation theory. Considerable effort has been
made in the last years to provide the matrix elements for all phenomenologically
relevant processes at next-to-leading order in the MC generators and to
interface them with the parton showers. The initial and final state parton
showers generate radiative corrections to the matrix elements by resumming
collinear or soft and collinear logarithms to leading logarthmic approximation
\footnote{In fact, modern parton showers also contain certain next-to-leading
logarithmic terms.}. These parts of
the event generators (i.e.\ matrix elements and parton showers) are nothing but
a faithful representation of perturbation theory with controlled systematic
accuracy. In contrast to this, the hadronisation is governed by
non-perturbative physics and is not accessible with perturbative methods. It is
instead simulated with the help of phenomenological models, the most commonly
used and best tested models are the Lund string
model and the cluster hadronisation. The model parameters are tuned to data (mainly from LEP) and assumed to
be universal.

\section{Jets in A+A}

In heavy ion collisions the hard matrix elements remain unchanged, due to the
high scale they involve. They thus happen on very short time and length scales
early in the reaction and are not affected by the formation of a system of
considerable density. The final state parton shower, on the other hand,
evolves over a considerably longer time scale and is most likely modified by
the formation of a dense and hot medium. There is, however, no general theory
for jet evolution in a dense medium, but only results for special cases (for
instance the single gluon radiation spectrum in the eikonal limit). The initial
state parton shower was found to be unmodified by the RHIC experiments. The
hadronisation, finally, is likely to experience modifications, but due to its
non-perturbative nature there is only very little theoretical guidance. 

\smallskip

The parton shower as introduced in section~\ref{sec_vacjets} is formulated in
momentum space. The theoretical description of jet evolution in a medium
requires a simultaneous formulation in momentum and configuration space. This
can be achieved by estimating the time scale for gluon radiation from the
uncertainty principle:
\begin{equation}
 \tau_\text{vac} \approx \frac{\omega}{\kt^2} \approx \frac{1}{\sqrt{t}}\cdot
\frac{E}{\sqrt{t}}
\end{equation}
The first expression is simply the inverse transverse energy of the radiated
gluon, while the second is the lifetime $1/\sqrt{t}$ of an unstable state
decaying into two partons characterised by a scale $t$ boosted to the
laboratory frame. The two arguments are parametrically equivalent. Apparently
the formation or decoherence of energetic fragments in the vacuum parton shower
is delayed by time dilation to times typically larger than the medium
length.

In the case of medium induced radiation the transverse momentum comes
predominantly from scattering in the medium: $\kt^2 = \qhat \tau_\text{med}$.
One thus finds for the lifetime of medium induced radiation
\begin{equation}
 \tau_\text{med} \approx \frac{\omega}{\kt^2} \approx \frac{\omega}{\qhat
\tau_\text{med}} \quad \Rightarrow
\quad \tau_\text{med} \approx \sqrt{\frac{\omega}{\qhat}}
\end{equation}
This means that soft emissions decohere first and at large angles. 

At face value these qualitative estimates indicate that jets emerging from
heavy ion collisions should have a hard core that fragments in vacuum
accompanied by soft medium induced radiation at large
angles\cite{CasalderreySolana:2010eh}. This is in
qualitative agreement with early jet measurements by ATLAS\cite{Aad:2010bu} and
CMS\cite{Chatrchyan:2011sx} and can be further clarified and quantified by
measurements of
intra-jet distributions (for instance fragmentation functions). 

\bigskip

The currently available MC models for jet quenching follow rather different
approaches. They shall only be briefly mentioned here, for details the reader
is referred to the original publications. 

\begin{description}
 \item[HIJING] introduces a medium induced parton splitting process,
collisional energy loss is neglected\cite{Wang:1991hta,Deng:2010mv}
 \item[HYDJET++/PYQUEN] simulates radiative energy loss by sampling a BDMPS
gluon spectrum and includes perturbative elastic
scattering\cite{Lokhtin:2008xi,Lokhtin:2005px}
 \item[JEWEL] aims for a unified description of all scattering processes in
terms of matrix elements and parton showers (work in
progress)\cite{Zapp:2008gi,Zapp:2008af}
 \item[Q-PYTHIA/Q-HERWIG] adds a term derived from BDMPS to the splitting
function and thus contains no elastic
scattering\cite{Armesto:2009fj,Armesto:2009ab}
 \item[YaJEM] assumes that medium interations increase the virtuality of
partons in the parton shower (leading to enhanced radiation) and can subtract
energy and momentum to simulated collisional energy
loss\cite{Renk:2008pp,Renk:2009nz}
 \item[MARTINI] is based on the AMY transition rates and also contains a
transition rate for elastic scattering\cite{Schenke:2009gb,Schenke:2011rd}
\end{description}

\bigskip

Going from single-inclusive to jet observables raises conceptual issues, that
will be discussed in the following.

\subsection{Non-eikonal kinematics}

The analytical calculations of radiative energy loss operate in the eikonal
approximation where
\begin{equation}
 E \gg \omega \gg \kt,\, \qt \gg \Lambda_\text{QCD} \,.
\end{equation}
Consequently, energy and momentum are not conserved and the scattering centres
are recoilless, i.e.\ there is no collisional energy loss. In contrast to this,
phenomenology at RHIC and LHC requires that all quantities in the above
equation can be of the same order, but are subject to energy-momentum
conservation. There are thus large uncertainties in the predictions of
analytical calculations due to kinematical ambiguities.

\smallskip

In general, i.e.\ non-eikonal, kinematics phase space restrictions due to
energy-momentum conservation have to be taken into account. This is not
necessarily a small effect, but can have rather dramatic effects on the gluon
spectrum. Furthermore, the scattering centres become dynamical and recoil
against the jet giving rise to collisional energy loss (also in inelastic
interactions) and induced radiation off the scattering centre. This already
hints at an additional complication, namely that elastic and inelastic
scatterings cannot be unambiguously separated from each other. They are two
possible outcomes (depending on a definition) of the same process rather than different processes. There is
also no clear separation between vacuum and medium induced radiation any more. 

\medskip

Incorporating exact energy-momentum conservation is for MC models typically
straight forward. The ambiguity between elastic and inelastic scattering, on
the other hand, requires a unified description of both processes, which is
currently under development. Dynamical scattering centres are problematic for
models based on effective descriptions for the medium interactions but are less
challenging for models incorporating microscopical interaction models.
Therefore, collisional energy loss is typically either neglected or added as a
separate process. Radiation off the scattering centres requires model dependent
assumptions and has not been included yet, but first steps are being taken.
Concerning the ambiguity between vacuum and medium induced radiation the models
either build on an unified description and don't distinguish at all between the
two, or they assume a complete factorisation where all vacuum radiation happens
first and the medium induced emissions follow afterwards.

\subsection{Multiple gluon emission and LPM-effect}

The analytical models only compute single gluon radiation, which is then
iterated probabilistically. This leads to complications due to the
non-conservation of energy and momentum in the eikonal limit. Concerning the
interplay between multiple gluon emissions first theoretical progress was made
recently through the calculation of gluon radiation off a colour
dipole\cite{MehtarTani:2011jw,CasalderreySolana:2011rz}.
The results indicate, that interference between emissions off the two legs
occurs only in restricted regions of phase space so that the dominant process
is independent radiation off both legs of the
dipole\cite{CasalderreySolana:2011rz}. 

\smallskip

Another important effect is that radiated gluons themselves radiate requiring a
democratic treatment of all partons. This can have significant effects in
particular on $\kt$- and angular distributions. This also means that the energy
loss is not a meaningful quantity any more and that rather the entire
fragmentation pattern has to be considered. Energy-momentum conservation is
already important for single gluon radiation, for the description of
multi-parton final states it is crucial. 

These considerations allow for the conclusion, that theories without democratic
radiation are not suitable for jet phenomenology.

\medskip

Treating the non-Abelian Landau-Pomerantchuk-Migdal-effect, which is a
quantum-mechanical interference, is notoriously difficult for MC models.
Therefore, most of them use effective descriptions of the single gluon
radiation process. However, in\cite{Zapp:2008af,Zapp:2011ya} a local and
probabilistic
formulation of the LPM-effect was derived, that will also be included in future
MC models.

\smallskip

The MC models also iterate single gluon radiation, which always
involves model dependent assumptions. The common assumption that all partons
radiate independently has received support from the recent results on gluon
radiation off colour dipoles. Democratic treatment is by construction easily
achieved (except for the scattering centre).

\subsection{$\kt$-broadening}

Understanding $\kt$-broadening is important as it affects the response of jet
finders to quenched jets. In the analytical models it is governed by Brownian
motion in the transverse space. It is, however, sensitive to energy-momentum
conservation, democratic multiple gluon radiation and contamination by
energetic recoils. 

\medskip

The assumptions in the MC models vary from collinear gluon emission to parton
shower kinematics and generally leave room for improvements of the microscopic
dynamics. 

\subsection{Recoils, medium modelling and background}

The propagation of highly energetic jets leads to modifications of the
medium in the vicinity of the jets, which is likely to survive into the
hadronic stage. A quantitative understanding of jet-induced modifications of
the medium is thus important for the experimental background subtraction. But
it is also interesting in its own right, as it for instance gives access to the
interplay between weakly and strongly coupled regimes. A satisfactory level of
understanding of the jet backreaction probably requires a unified description
of jet and medium evolution. While this is still some way to go first attempts
to characterise the reaction of the medium to jets have been made from the jet
quenching side by tracking the recoiling scattering centres\cite{Zapp:2008gi}
and from the
medium modelling side by solving hydrodynamics with source terms that describe
the energy and momentum deposition of a jet\cite{Neufeld:2010tz}.

\medskip

The fact that most MC models use hydrodynamic calculations as model for the
medium makes it difficult for them to quantify the backreaction. A conversion
of hydrodynamical results into a population of scattering centres is possible,
but involves model dependent assumptions. Even then, the information about the
distortion due to the passage of a jet cannot be propagated back to the
hydrodynamical calculation. Event-by-event hydrodynamics is a development that
may allow for a simultaneous solution in the future, but the problem of having
two vastly different approaches for the jet and the medium evolution remains.
This is partly a problem of very different scales, as the jets are described
in perturbation theory requiring a high scale, whereas the bulk medium
evolution is governed by non-perturbative processes at much lower scales.
Consequently, also the language is vastly different with the jets being
formulated in terms of individual partons while the medium is usually treated
using continuum dynamics. Bringing both regimes together is thus far from
trivial. 
Parton cascades that are built on a partonic language also for the medium have
so far not been able to consistently include the parton shower.

\subsection{Hadronisation}

Medium modifications to the hadronisation phase are likely and raise
conceptually both interesting and difficult questions. Due to the
non-perturbative nature of the problem there is at best very little theoretical
guidance. Therefore, it is commonly assumed that hadronisation happens at late
times and therefore in vacuum. This is a reasonable assumption for energetic
fragments due to the large boost factors, but not necessarily for soft and
semi-hard fargments. Even if one assumes hadronisation in vacuum there are
complications due to the medium: The hadronisation is sensitive to the colour
topology of the event, but strong interactions in a medium inevitably alter the
colour configuration. Furthermore, it is unclear how jet and medium
hadronisation
interplay, in particular in regions of phase space where soft fragments of the
jets overlap with relatively hard fluctuations or recoils from the medium. This
leads to potentially large uncertainties even in a factorised approach. 

\medskip

MC models assume -- like most other models -- that hadronisation happens in
vacuum. Some
allow for modifications of the colour toplogy, but with the exception of
\mbox{Q-HERWIG} they all rely on the Lund string model. As it is not clear how to
sytematically improve the existing approaches, the currenly most promising
strategy is to understand and quantify the systematics and uncertainties of the
models, for instance by varying assumptions about the colour topology, use of
different hadronisation models etc. 

Some of the implemented prescriptions have a potentially dangerous shortcoming
in that they are infra-red and/or collinear sensitive. This is a point that 
could be improved upon. 

Finally, there are alternative ideas (like pre-hadron formation) that should in
principle be suitable for a MC implementation. Comparing new ideas to the
traditional approaches could
perhaps also lead to new insights.

\section{Conclusions}

There is strong motivation from both the theoretical and experimental side for
studying jets in heavy ion collisions, although this implies a considerable
increase of complexity in the theoretical description. Among the conceptual
issues raised by the transition from single-inclusive quantities to jets
non-eikonal kinematics, multi gluon emission, $\kt$-broadening, jet-induced medium
modifications and hadronisation are the most important ones. MC generators are
powerful and versatile theory tools that allow to explore all these issues. They
are designed to describe jets on the basis of multi-particle final states and
allow to account dynamically for the interactions between the jet and the
medium. However, as long as there is no general theory of jet quenching, also
MC generators will have to rely on phenomenological models. It is to be
expected that there will be considerable progress in the next years, driven by
the jet data from RHIC and LHC and fruitful interaction between theorists and
experimentalists.  

\ack
The author would like to thank Urs~Wiedemann for many enlightening discussions.

\section*{References}


\end{document}